# Dimensionality of vortex matter in superconducting infinite-layer nickelates

D. Sanchez-Manzano[1,#], V. Humbert[1,#], D. Zhang[1], A. Gutiérrez-Llorente[1,2], M. Bibes[1], J. Santamaría[3], L. Iglesias[1,*], Javier E. Villegas[1,&]

[1]*Laboratoire Albert Fert, CNRS, Thales, Université Paris-Saclay, 91767 Palaiseau, France*

[2]*Escuela Superior de Ciencias Experimentales y Tecnología, Universidad Rey Juan Carlos, 28933 Madrid, Spain*

[3]*GFMC, Dpto. de Física de Materiales, Facultad de Ciencias Físicas, Universidad Complutense de Madrid, 28040 Madrid, Spain*

[#] contributed equally
* lucia.iglesias@cnrs-thales.fr
[&] javier.villegas@cnrs-thales.fr

Characterizing the dimensionality of the superconducting state in infinite-layer (IL) nickelates is essential for understanding its nature. Most studies have addressed this by examining the anisotropy of the upper critical fields. However, the dominance of Pauli paramagnetic effects over orbital effects complicates the interpretation of these experiments in terms of dimensionality. Here, we approach the question from a different perspective by mapping the vortex phase diagram. We show that superconducting $Pr_{0.8}Sr_{0.2}NiO_2$ thin films with low disorder exhibit a vortex liquid-to-glass transition of a *quasi*-two-dimensional (2D) nature. In contrast, increasing disorder drives a crossover into a *pure* 2D state. This demonstrates that pure bidimensionality is an extrinsic property, resulting from the decoupling of $NiO_2$ planes due to enhanced disorder. Our findings establish disorder as a key control parameter of superconductivity in IL nickelates and suggest that it resides within the $NiO_2$ planes, providing two fundamental insights for understanding these materials.

Since the discovery of superconductivity in infinite layer (IL) nickelates [1], many efforts have focused on finding analogies and differences with the cuprates [2,3], with which they share electron counts ($Ni^{1+}$:$3d^9$) and a square planar geometry in the basal plane. This comparison aims to explore potential shared pairing mechanisms [4], which may provide new insights into the longstanding problem of high-temperature superconductivity. Research has examined d-wave superconductivity [5–7], charge density waves [8–11], and strange-metal behavior [12], all hallmarks of cuprates. Meanwhile, key differences have emerged: unlike cuprates, whose parent compounds are antiferromagnetic insulators, undoped IL-nickelates exhibit bad-metal behavior or superconductivity [13], and the existence of long-range antiferromagnetism remains debated [8,14–16].

Superconductivity in IL-nickelates has only been observed in ultrathin films with a layered, anisotropic structure where $NiO_2$ planes are separated by rare-earth hole dopants (Nd, Pr, La, Sr, Ca) [1]. This structure has led to extensive studies of the superconducting phase anisotropy and dimensionality by measuring upper critical fields $H_{C2}$ along different crystallographic directions [17–24]. Some reports describe highly anisotropic behavior [20–23], while others find nearly isotropic [17–19] superconductivity, with a strong dependence on the rare-earth ion [21] and doping level [18]. A major challenge for interpreting these results is the competition between orbital-limited and Pauli-limited superconductivity [17], which obscures the connection between $H_{C2}$ anisotropy and dimensionality.

Despite being a key probe of the superconducting state, vortex physics in IL-nickelates has received limited attention [25,26], and their vortex phase diagram remains largely unexplored. In cuprates, thermal effects, anisotropy, vortex elasticity, and disorder create a rich phase diagram [27]. Clean single crystals exhibit a zero-resistance vortex-solid phase at low temperatures, separated by a first-order melting transition from a dissipative vortex-liquid [28].

In thin films, disorder transforms this into a continuous liquid-to-vortex-glass (VG) transition at a glass temperature $T_g$, below which dissipation vanishes despite lacking long-range translational order [27,29]. Whether IL-nickelate films exhibit similar behavior is unknown. This is critical because the VG transition is highly sensitive to the dimensionality of the superconducting state: the VG correlation length $\zeta_{VG}$ must diverge when approaching the glass phase, but this divergence is suppressed in reduced-dimensionality systems. In strictly 2D systems, divergence occurs only in-plane, resulting in $T_g = 0$ K and no zero-resistance state at finite temperature [30].

Here, we investigate vortex dynamics to determine the dimensionality of vortex matter in IL $Pr_{0.8}Sr_{0.2}NiO_2$ (PSNO) films, using samples with a wide range of superconducting critical temperatures $T_C$ and normal-state resistivities, $\rho$. Films with the highest $T_C$ and lowest $\rho$ show quasi-2D behavior, characterized by finite VG correlations and superconducting coupling perpendicular to the $NiO_2$ planes. Increasing disorder (higher $\rho$) drives a crossover to purely 2D vortex behavior, with vanishing VG correlations perpendicular to the planes. This indicates that superconductivity in IL-PSNO resides primarily in the $NiO_2$ planes, whose coupling is suppressed by disorder, leading to the dimensional crossover.

Perovskite $Pr_{0.8}Sr_{0.2}NiO_3$ (PV-PSNO) thin films were epitaxially grown on (001)-oriented $SrTiO_3$ (STO) substrates by Pulsed Laser Deposition. Two reduction methods produced the IL-PSNO phase. For sample #1, a ~ 2 nm STO capping layer was deposited prior to ex situ $CaH_2$ reduction to stabilize the IL phase [8,31,32]. For samples #2–#4, we stabilized the IL phase via Al-assisted reduction using a ~2.5 nm thick Al overlayer [33]. Figure 1 (a) displays typical X-ray θ-2θ scans around the (001) reflection, before and after reduction. Upon reduction, the (002) diffraction peak shifts from 48.4° (PV-PSNO) to 54.8° (IL-PSNO) as expected [32,34,35]. The films thicknesses $d$ ~7–8 nm were determined by low-angle X-Ray reflectivity (Fig. S1; Table

S1). We patterned standard 4-probe bridges using UV lithography and ion etching for samples #2–#4 (Fig. S2).

Figures 1(b)–(c) display typical resistance *vs.* temperature at zero field and under applied fields. Fig. 1(d) shows that the zero-field $T_C$ decreases systematically with increasing $\rho$ (measured above the transition onset), consistent with prior observations [33] (see also Fig. S2). The disorder responsible for increased $\rho$ and suppressed superconductivity likely involves oxygen-related defects, including residual apical oxygen and interstitials in the Pr/Sr spacer layers [7,33,36]

The superconducting transition broadens under applied fields, more strongly for fields perpendicular to the film plane [Fig. 1(c)] than for fields parallel to the $NiO_2$ (or *ab*) planes [Fig. 1(b)]. This anisotropy resembles that in $La_{0.8}Sr_{0.2}NiO_2$ (IL-LSNO) [18,20,23] and exceeds that in $Nd_{0.775}Sr_{0.225}NiO_2$ (IL-NSNO) [17,24]. $R(H)$ qualitatively behaves as expected [17] for both perpendicular [Fig. 1(f)] and in-plane fields (Fig. S3). Using the criterion $R(\mathrm{H_{c2}}) = 0.5R_N$ we calculated the upper critical fields [Fig. 1(g)]. For all samples, $\mathrm{H_{c2,\perp}}$ shows linear behavior, while $\mathrm{H_{c2,\parallel}}$ (parallel to the $NiO_2$ plane) follows a square-root dependence, as the Ginzburg-Landau (GL) approximation predicts for films of thickness $d$ [37]:

(1) $$H_{C2,\perp}(T) = \frac{\phi_0}{2\pi\xi_{ab}^2(0)}(1 - T/T_C)$$

(2) $$H_{C2,||}(T) = \frac{\sqrt{12}\phi_0}{2\pi d\xi_{ab}(0)}(1 - T/T_C)^{0.5}$$

From $H_{C2,\perp}(T)$ and Eq. 1, we derived the in-plane coherence length $\xi_{ab}$ in the $NiO_2$ plane [see inset in Fig1(g)]. The zero-temperature values $\xi_{ab}(0)$ [Fig. 1(e)] align with IL-NSNO reports [1,17,38,39]. As in Ref. [17], $H_{C2,||}(T)$ is lower than Eq. 2 predicts using measured $d$ and $\xi_{ab}(0)$ [Fig. 1(g)], suggesting Pauli-limited superconductivity [17,40]. In this scenario, $H_{C2,||}$ is unrelated to $d$, $\xi_{ab}$ and $\xi_c$. It is instead determined by $H_{C2,||} = H_p = \Delta_0(T)/\sqrt{2}\mu$ ,

with $\Delta_0(T) = \Delta_0(0)(1 - T/T_C)^{0.5}$ the zero-field superconducting gap and $\mu$ the effective in-plane electronic magnetic moment. Assuming BCS behavior $\Delta_0(0) = 1.76k_B T_C$, we find $\mu\sim 0.5\mu_B - 0.65\mu_B$ for samples #1–#4 (Supplementary Table S1), similar to IL-PSNO (see Fig. 4e in [24]), $La0_{.8}Ca_{0.2}NiO_2$ (see Fig. 2d in [18]) and IL-LSNO (table S1 in [20], but contrasting with enhanced $\mu\sim 2.4\mu_B$ in IL-NSNO [17]. Possible origins for this diverse phenomenology –rare-earth magnetism [24], triplet superconductivity [18], or spin-orbit effects [41,42]– remain debated but are beyond this work's scope. The key conclusion is that $H_{C2,||}$'s independence from sample dimensions and coherence lengths means that angular-dependent critical fields and magnetoresistance [21] should not be interpreted via vortex dimensionality.

Figure 2 shows Arrhenius plots of $R(T)$ for fields perpendicular to the film plane. For all samples and fields, $\log(R)$ vs. $1000/T$ is linear over much of the resistive transition, indicating Thermally Activated Flux-Flow (TAFF) regime [27] with $R \propto \exp[-U(H)/k_B T]$. Fig. 2 (b) and (c) display the vortex activation energy $U(H)$ calculated from Arrhenius plots. At high fields, where collective vortex pinning expectedly dominates [27], $U(H) \propto -\log(H)$, typical of low-dimensional vortex matter [27], and has been observed in a variety of systems including the most anisotropic cuprates [43–45]. This contrasts with $U(H) \propto H^{-\alpha}$ observed in 3D superconductors (e.g. moderately anisotropic $YBa_2Cu_3O_7$ where $0.5 < \alpha < 1$ [46]), as well as in IL-NSNO where $\alpha > 5$ [17]. For samples #1–#2, $U(H)$ weakens at low fields (<1 T) as observed earlier [17], suggesting a crossover to single-vortex pinning, as expected with strong disorder [27]. Figure 2 also shows TAFF deviations at low temperatures, where R exceeds extrapolated Arrhenius trends below a departure temperature $T_d$ (and resistance level $R_d$), which decreases (increases) with field (Fig. S7). This behavior, observed only in samples #1–#2, suggests a crossover from TAFF to quantum vortex motion [47], as seen in 2D systems [43,48–50].

Figures 3(a) and 3(c) show isothermal $V(I)$ curves in fixed magnetic fields, for samples #4 and #2. At high temperatures, the response is ohmic across the current range. Upon cooling, the ohmic range narrows: nonlinearity emerges above a current $I_{nl}$ [see Fig. 3(a)] that decreases with temperature, along with the ohmic resistance $R_{lin} = \lim_{I \to 0} V/I$. This matches expectations for a continuous vortex-liquid to VG transition [29,51–57], where below the glass transition temperature $T_g$, the $V(I)$ curvature changes from upward to downward concavity at low currents, yielding $R \to 0$. That is the case for the data in Fig 3 (a), whose derivatives $d(\log V)/d(\log I)$ vs $\log I$ [inset of Fig. 3(b)] below 7.75 K and above 8 K show opposite curvature in the low-current limit, implying 7.75 K< $T_g < 8$ K [58]. In contrast, in Fig. 3(c) all curves remain concave upward [inset of Fig. 3(d)], implying either a finite $T_g$ <1.8 K (the base temperature of the experiment) or pure 2D behavior with $T_g = 0$ [30,59]. As shown below, the latter is confirmed by the quantitative analysis of the derivatives, as well as by the collapse of the $V(I)$ according to VG scaling theory.

For 3D or quasi-2D (q2D) systems, the VG theory [51,60,61] predicts the scaling ansatz:

$$(V/I)\left|1 - T/T_g\right|^{\nu(D-2-z)} = g_{\pm}\left((I/T)\left|1 - T/T_g\right|^{\nu(1-D)}\right) \quad (3)$$

where $g_{\pm}$ is a (field-dependent) universal function for all $T$ above (+) and below (-) $T_g$, $D = 3$ for 3D and $D = 2$ for q2D , and $z$ and $\nu$ are the static and dynamic critical exponents which theory predicts $4 < z < 6$ and $1 < \nu < 2$ [51]. The critical isotherm $-V(I)$ at $T_g-$ fulfills $V \propto I^{\alpha+1}$ with $\alpha = (z - D + 2)/(D - 1)$ in the low current limit [51,60,61]. Consequently, $d(\log V)/d(\log I)$ around $T_g$ must satisfy 5< $\alpha + 1$ <7 for $D$=2 and 2.5< $\alpha + 1$ <3.5 for $D$=3 [indicated by grey shading in the insets of 3(b) and 3(d)]. Thus, the inset of Fig. 3 (b) allows the unequivocal determination of $D$=2, 7.75 K< $T_g < 8$ K and $z = \alpha \sim$ 4.9 (horizontal line). Consistently, an excellent scaling of the the corresponding $V(I)$ curves is obtained by slitghtly refining those paremeters and using $\nu = 1.55$ (within the range expected from theory), see Fig. 3 (b). In contrast, the same analysis of the inset of Fig. 3 (c) shows that no finite $T_g$ exists, since all the isotherms have upwards concavity, both below and above the derivative ranges expected for 3D and q2D (shaded grey). Consistently, for this dataset an excellent scaling is obtained using the pure 2D asnsatz [30,54,57,59]:

$$(4) \qquad \frac{V}{I} \cdot e^{\left(\frac{T_0}{T}\right)^p} = f\left(\frac{I}{T^{1+\nu_{2D}}}\right)$$

where $f$ is a (field-dependent) universal function, $T_0$ is an energy scale related to the vortex energy, the critical exponent $\nu_{2D}$ =2, and $p$ reflects the dominant transport mechanism, with $p \geq 1$ for thermal activation $p \approx 0.7$ for quantum vortex tunneling [61]. Consistent with the TAFF observed in Fig. 2, we achieve an excellent collapse [Fig. 3(d)] with $p = 1$ and $T_0$ as the only free parameter (with a ~10% error, Fig. S4).

We applied this analysis to all datasets (samples #1–#4). Sample #4 (lowest $\rho$) behaves as q2D for 0.1 T$< \mu_0 H < 9\,T$ [Figs. 3 (b), S6 and S9]. In contrast, samples #1–#2 (higher $\rho$, lower $T_c$) show pure 2D behavior (Fig. 3(d), and S8). Sample #3 also shows pure 2D behavior, though q2D scaling is possible for $\mu_0 H \leq$1 T but only with $z$>10 – well beyond the range predicted by theory (Figs. S5 and S6). It appears that sample #3, with $T_C$ slightly below sample #4 but ~double $\rho$, sits at the boundary between q2D and 2D behavior for $\mu_0 H \leq 1$ T, shifting to pure 2D at higher fields. For all pure 2D cases, $T_0(H)$ from collapses [Fig. 3(c), inset] closely matches $U(H)$ in Figs. 2(b)–(c), further confirming the $p = 1$ consistency.

In summary, the $V(I)$ analysis indicates vortex matter in IL-PSNO undergoes a crossover from q2D to purely 2D behavior, driven by increased normal-state resistivity. As discussed below, this dimensional crossover can be directly linked to the suppression of interplane vortex correlations, which reflects in characteristic vortex length scales perpendicular to the $NiO_2$ planes.

For sample #4 (q2D for all fields), the VG correlation length is finite along $NiO_2$ planes ($\zeta_{VG.\parallel}$) and perpendicular to them ( $\zeta_{VG.\perp}$). We can estimate an upper limit for $\zeta_{VG.\perp}$ from $I_{nl}$, where the Lorentz force work equals thermal energy, i.e. $j_{nl}\phi_0\zeta_{VG,\perp}\zeta_{GV.\parallel} \approx k_BT$ [51,63] ($\phi_0$ the flux quantum). We can safely assume [64] that for the highest $T$ isotherm $\zeta_{VG,\parallel} > 2a_0 \approx 2\sqrt{\phi_0/\mu_0 H}$ (with $a_0$ the intervortex distance), and thus that $\zeta_{VG,\perp} < k_BT\ /2j_{nl}\sqrt{\phi_0^3/\mu_0 H}$. Figure 3(a) estimates show $\zeta_{VG,\perp}$ ~ $d$ (film thickness, horizontal line), consistent with q2D scaling. Samples #1–#3's pure 2D behavior implies $\zeta_{VG.\perp} = 0$ [65,66]. Here we can estimate the vortex length $l$ from $T_0$ using [54,59] $k_BT_0 \approx \varepsilon_0 l$, with $\varepsilon_0 = \hbar^2\rho_S/m = \hbar^2/\mu_0 e^2\lambda_L^2$ the vortex core energy per unit length. $\rho_S$ and $m$ are the Cooper-pair density and mass, and the penetration length $\lambda_L = \sqrt{m/\mu_0\rho_S e^2}$ ~ 1 μm in IL-PSNO [6]. Figure 3(a) shows that $l$ gradually decreases with $\rho$, becoming significantly shorter than the film thickness and eventually comparable to the $NiO_2$ inter-planes distance.

The conclusion from the above is that IL-PSNO is not intrinsically 2D. Instead, disorder-driven resistivity increases cause a dimensional crossover by shortening the vortex length. This suggests superconductivity resides in $NiO_2$ planes with weak coupling that is highly susceptible to disorder (likely oxygen-related defects as discussed above [7,33,36]). The emerging picture is that coherence length $\xi_c$ is shorter than the interplane distance −consistent with Pauli-limited behavior− and that coupling is governed solely by Josephson and/or vortex magnetostatic interactions [27]. In this scenario, vortex pinning and in-plane vortex interactions overcome interplane coupling, yielding uncorrelated "pancake vortices" [67] of thickness $l < d_{PSNO}$, and a vanishing $\zeta_{VG,\perp}$. This resembles the case of $Tl_2Ba_2CaCu_2O_8$ [59] and oxygen-depleted $YBa_2Cu_3O_{7-\delta}$ [54], where q2D behavior (finite $T_g$) at low fields crosses over to pure 2D one ($T_g = 0$) above ~1 T [68]. For disordered IL-PSNO, this occurs at ~0.1 T. Two factors explain this fast crossover despite the short $NiO_2$ planes separation. First, a relatively long $\lambda_L$~ 1 µm [6] (~100 nm in $YBa_2Cu_3O_7$ [69] and ~200 nm in $Tl_2Ba_2CaCu_2O_8$ [70]), enhancing in-plane vortex interactions at low fields [27]. Second, the possible Pr field-induced magnetism [24] may further suppress interplane coupling.

In summary, we studied IL-PSNO's vortex matter dimensionality. While the cleanest films are quasi-2D, disorder drives a crossover to pure 2D behavior. Results suggest $NiO_2$-plane-confined superconductivity with weak interplane coupling. We believe this conclusion constitutes an important ingredient for understanding the nature of superconductivity in the IL-nickelates. The revealed dominance of disorder in driving the dimensional crossover may be relevant beyond the nickelates. For example, some cuprates show similar crossovers via oxygen stoichiometry changes [64], though the mechanisms remain unknown. Our findings suggest that disorder inherently linked to stoichiometric changes [71] may be a key factor for these materials also.

**Acknowledgements**

We acknowledge funding from the French ANR (ANR-22-CE24-0009-01 "SEEDS", ANR22-CE30-00020-01 "SUPERFAST", ANR-22-EXSP-0007 PEPR SPIN "SPINMAT"), the EU's EIC Pathfinder grant 101130224 "JOSEPHINE", and COST action "SUPERQUMAP". A.G.L. acknowledges financial support through a research grant from the Next Generation EU plan 2021, European Union.

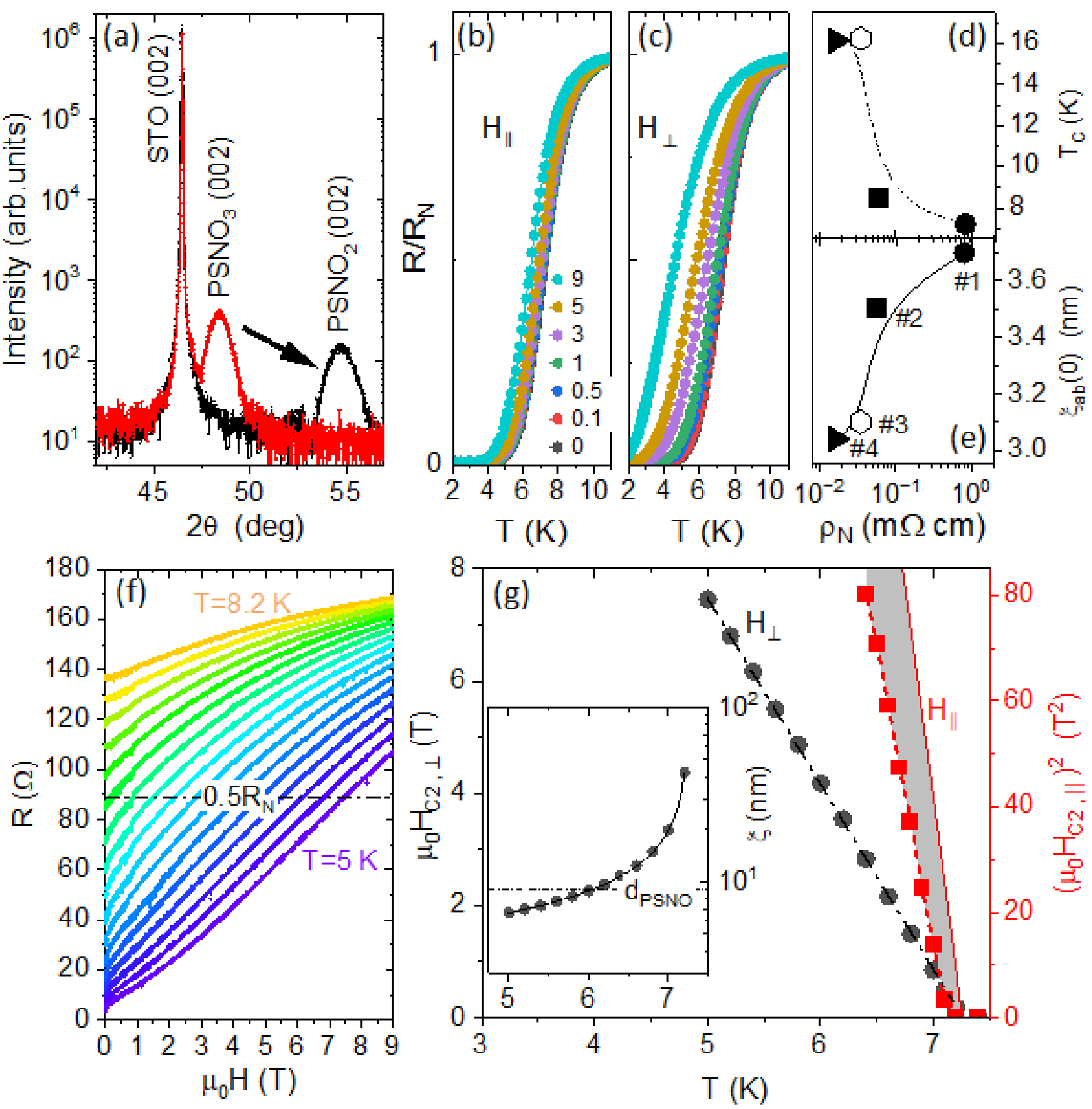

**Figure 1** (a) X-ray diffraction θ-2θ scan (Cu K-$\alpha$ radiation) of sample #1, measured as-grown (red) and after reduction (black). (b) Normalized resistance as a function of temperature $T$ for sample #1, measured with $I = 10$ μA and magnetic field $H$ (see legend in T) applied parallel and (c) perpendicular to the $NiO_2$ planes. (d) Superconducting transition temperature $T_C$, defined by the criterion $R = 0.5R_N$, and (e) in-plane coherence length as a function of the normal-state resistivity (measured above the transition onset) for samples #1-#4. (f) Resistance vs perpendicular $H$ for sample #1, measured in the temperature range 5 K $< T <$ 8.2 K in steps of 0.2 K. (g) Upper critical field $H_{c2,\perp}$ (left axis) and $\left(H_{c2,\parallel}\right)^2$ (right axis) as a function of $T$. Lines are guides to the eye. The grey shaded region highlights the difference between $\left(H_{c2,\parallel}\right)^2$ and the expectation from Eq. 2 (red solid line). Inset: In-plane coherence length $\xi(ab)$ vs. $T$. The horizontal dashed line indicates the film thickness $d_{PSNO}$.

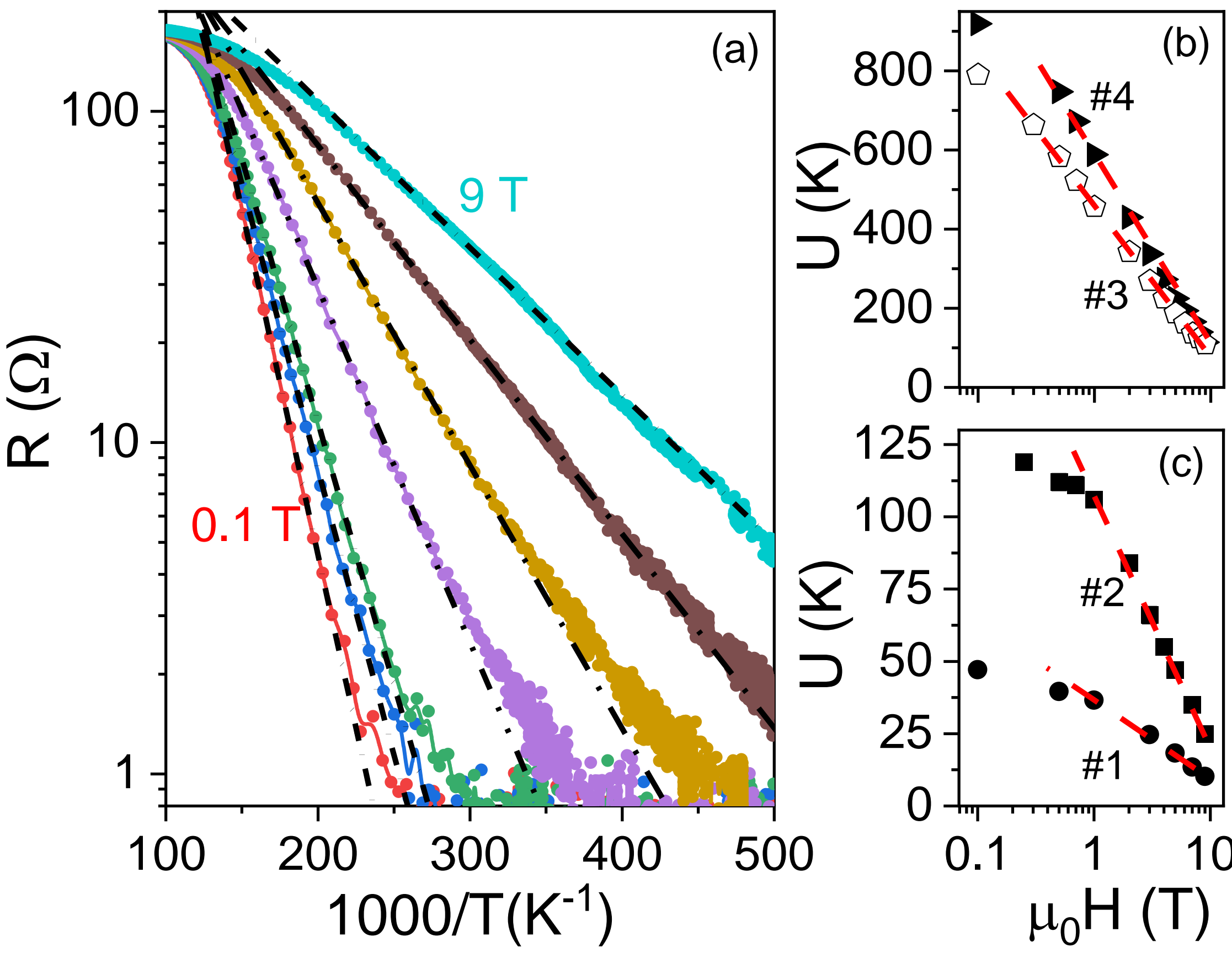

**Figure 2.** Resistance vs $1000/T$ for sample #1 measured under perpendicular magnetic fields (from left to right): 0.1 T, 0.5 T, 1 T, 3 T, 5 T, 7 T and 9 T. Black dashed lines are linear fits. (b) and (c): activation energy $U$ vs. $H$ for samples #1-#4. The dashed lines are fits for $\mu_0 H >$ 0.5 T.

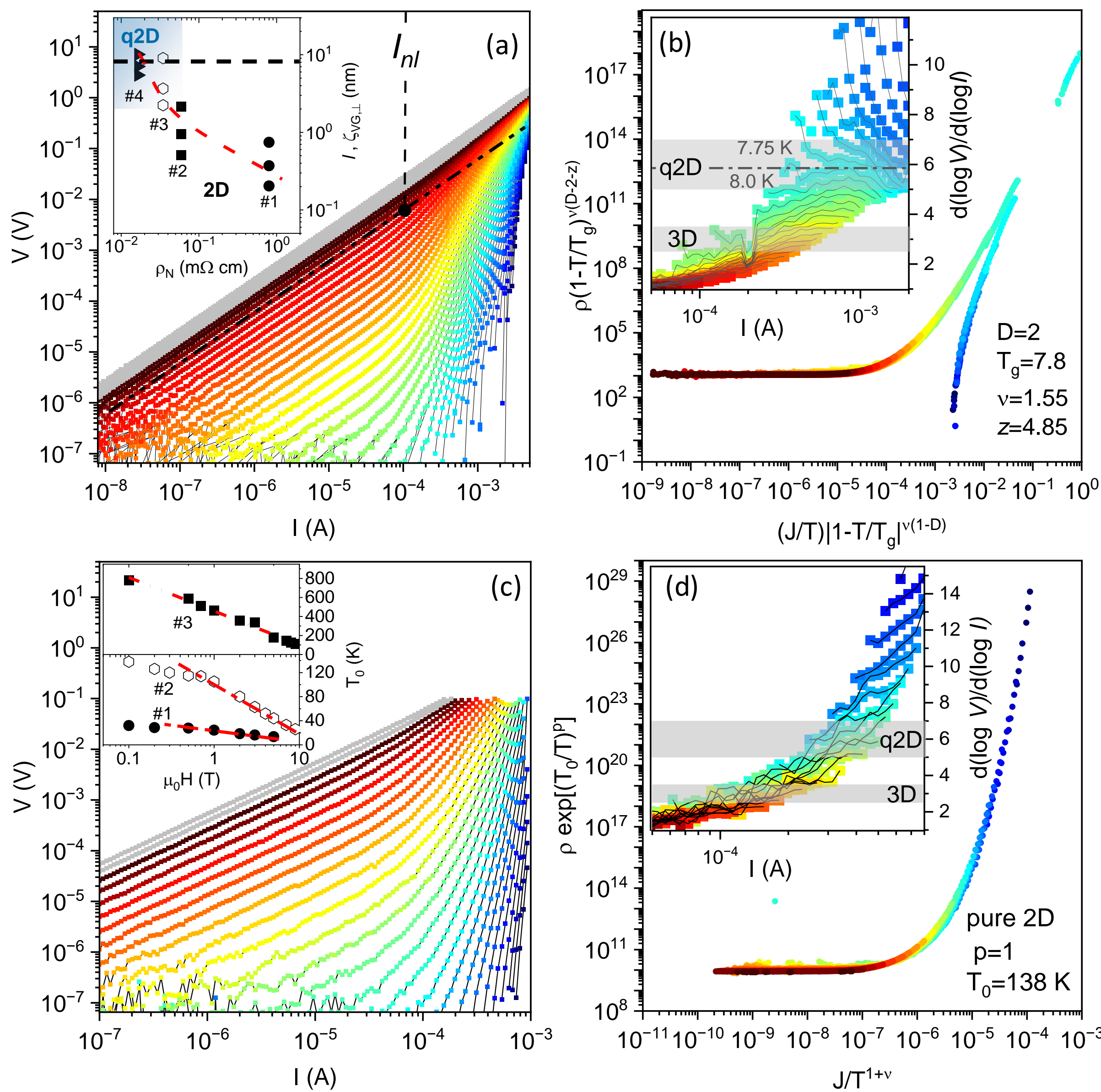

**Figure 3** (a) For sample #4: $V(I)$ isotherms measured in the range 4.25 < T < 16 K in ~ 0.25 K steps, for $\mu_0 H_\perp$=4 T. Inset: vortex length $l$ estimated from T$_0$ (samples #1-#3) and $\zeta_{VG,\perp}$ estimated from $j_{nl}$ (sample #1) for magnetic fields of 1, 5, and 9T (top to bottom). The line is a guide to the eye. (b) Scaling of the isotherms in (a) following a quasi-2D VG model (Eq. 3). Isotherms for which $R_{lin} > 0.5R_N$ (in grey) are excluded, in agreement with the $H_{c2}$ criterion. Inset: derivative of the curves in (a), used to determine T$_g$ and $z$. The shaded grey areas indicate the expected range of slopes for the critical isotherm within the q2D/3D vortex-glass theory. (c) For sample #2: $V(I)$ isotherms with 2 < T <8.5 K in ~ 0.25 K steps, for $\mu_0 H_\perp$= 0.1 T. Inset: T$_0$ (H) for purely 2D samples #1-#3. Dashed lines are linear fits for $\mu_0 H_\perp$> 0.5 T. (d) Scaling of the isotherms in (c) following a purely 2D model (Eq. 4). Inset: derivative of the curves in (c).

**Supplemental Material for**

# Dimensionality of vortex matter in superconducting infinite-layer nickelates

D. Sanchez-Manzano[1], V. Humbert[1], D. Zhang[1], A. Gutiérrez-Llorente[1,2], M. Bibes[1], J. Santamaría[3], L. Iglesias[1,*], Javier E. Villegas[1,&]

[1]*Laboratoire Albert Fert, CNRS, Thales, Université Paris-Saclay, 91767 Palaiseau, France*
[2]*Escuela Superior de Ciencias Experimentales y Tecnología, Universidad Rey Juan Carlos, 28933 Madrid, Spain*
[3]*GFMC, Dpto. de Física de Materiales, Facultad de Ciencias Físicas, Universidad Complutense de Madrid, 28040 Madrid, Spain*

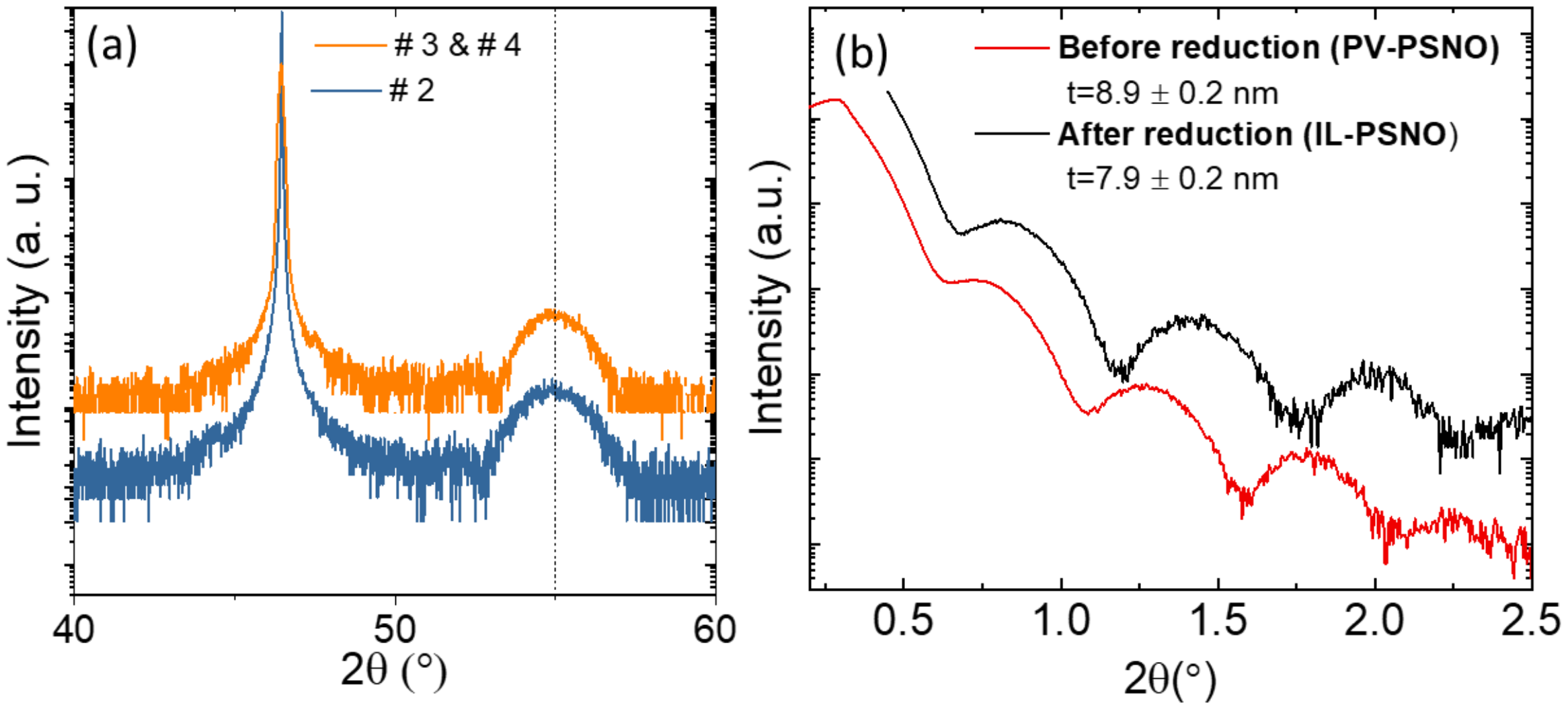

**Figure S1: (a)** XRD pattern of samples #2-#4. The sharp peak corresponds to the (002) reflection of the $SrTiO_3$ substrate, while the broader peak arises from the (002) reflection of the infinite-layer phase of the PSNO film. Notice that the absence of the (002) reflection of the perovskite phase of PSNO indicates complete reduction of the PSNO film. (b) Representative x-ray reflectivity (XRR) curves of a PSNO measured before and after reduction. In all cases, the film thickness was extracted from Fourier analysis of the Kiessig fringes and independently validated by in situ reflection high-energy electron diffraction (RHEED) intensity oscillations recorded during the growth of each parent perovskite film.

**Supplementary Table 1**

| **Sample** | ***d*** **(nm)** | $\boldsymbol{\mu}$ **(**$\boldsymbol{\mu_B}$**)** |
|---|---|---|
| #1 | 7.9±0.2 | 0.5 |
| #2 | 6.9±0.2 | 0.58 |
| #3 | 7.1±0.2 | 0.5 |
| #4 | 7.1±0.2 | 0.65 |

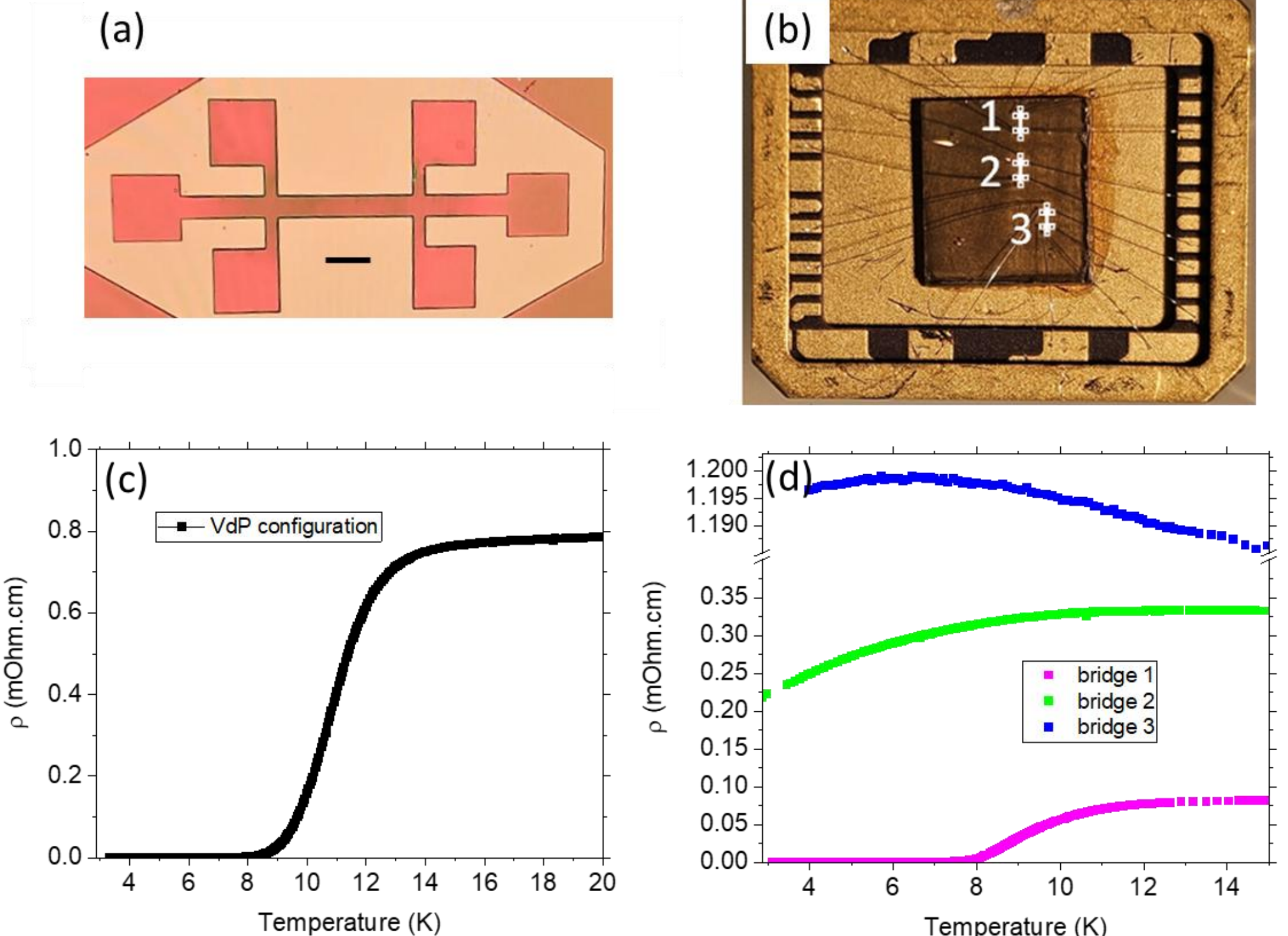

**Figure S2: (a)** Optical microscopy image of a bridge after etching the IL-PSNO by ion beam etching (photoresist still on top). The scale bar is $100\mu$m. (b) Optical microscopy image of an ~5 mm x 5 mm STO substrate on the cryostat's sample holder. The STO substrate contains 12 lithographed IL-PSNO bridges. Bridges 1, 2, and 3 (their location on the substrate is indicated by the white markers) are wide bonded to the sample holder. (c) Resistivity vs. temperature measured before bridges are lithographed, on the plain IL-PSNO film on STO. The resistivity is calculated from the resistance measured in four-probe configuration, using the Van der Pauw (VdP) method. (d) Resistivity vs. field of the bridges 1, 2 and 3 lithographed in the same film measured in (c). Superconductivity is only found in bridge 3, the two other bridges show very different normal-state resistivity (here it is calculated from the measured resistance considering the bridge dimensions). This shows that the IL-PSNO films show inhomogeneities on the mm scale, and that caution must be taken when extracting certain parameter from measurements in non-lithographed samples. By selecting the best lithographed bridges from series of films, we could observe some of the highest $T_C$/lowest resistivity in the literature.

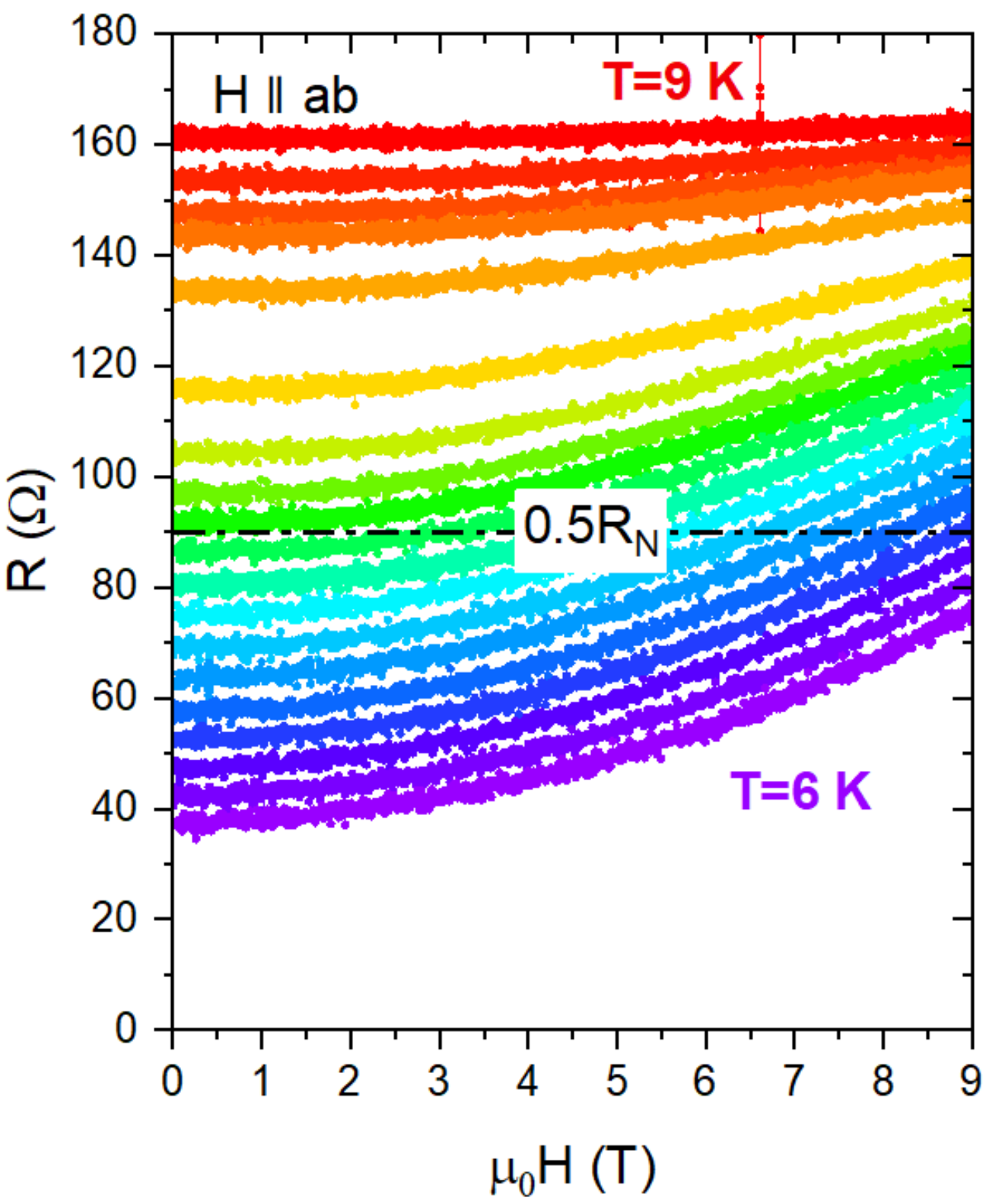

**Figure S3:** Resistance vs. H applied along the *ab* axis for sample #1, for 6 K $< T <$ 9 K. The horizontal line indicates the resistance criterion used to determine the upper critical field $H_{c2}$..

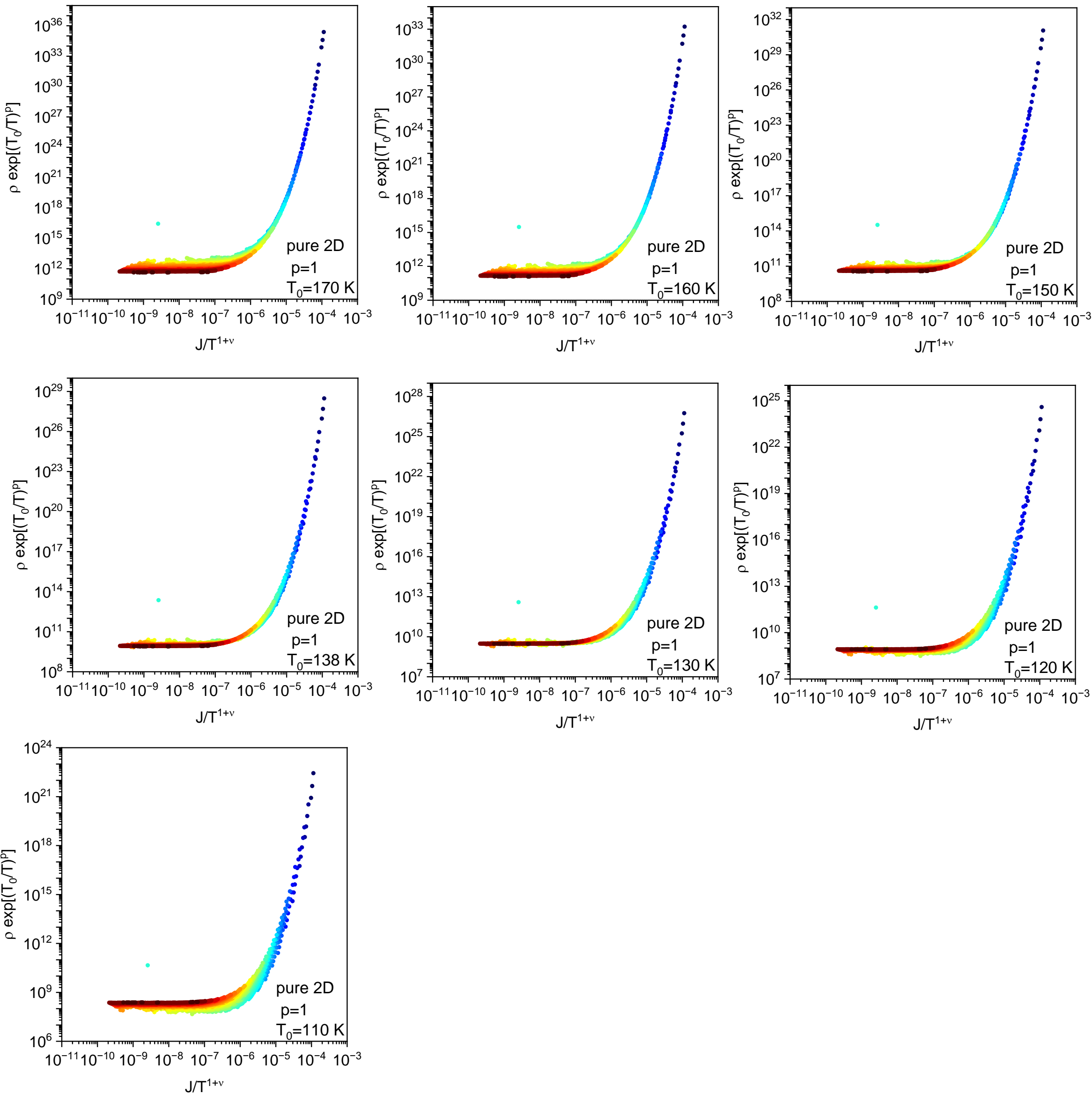

**Figure S4:** Scaling of the $V(I)$ dataset shown in Fig. 3 (c), using different values of the scaling parameter $T_0$. When $T_0$ is around 10% off the optimum $T_0 = 138\ K$, a sensible degradation of the data collapse is observed. The degradation becomes much stronger when $T_0$ is 20% off. This gives a sense of how critical the scaling parameter is, and how precisely it can be determined.

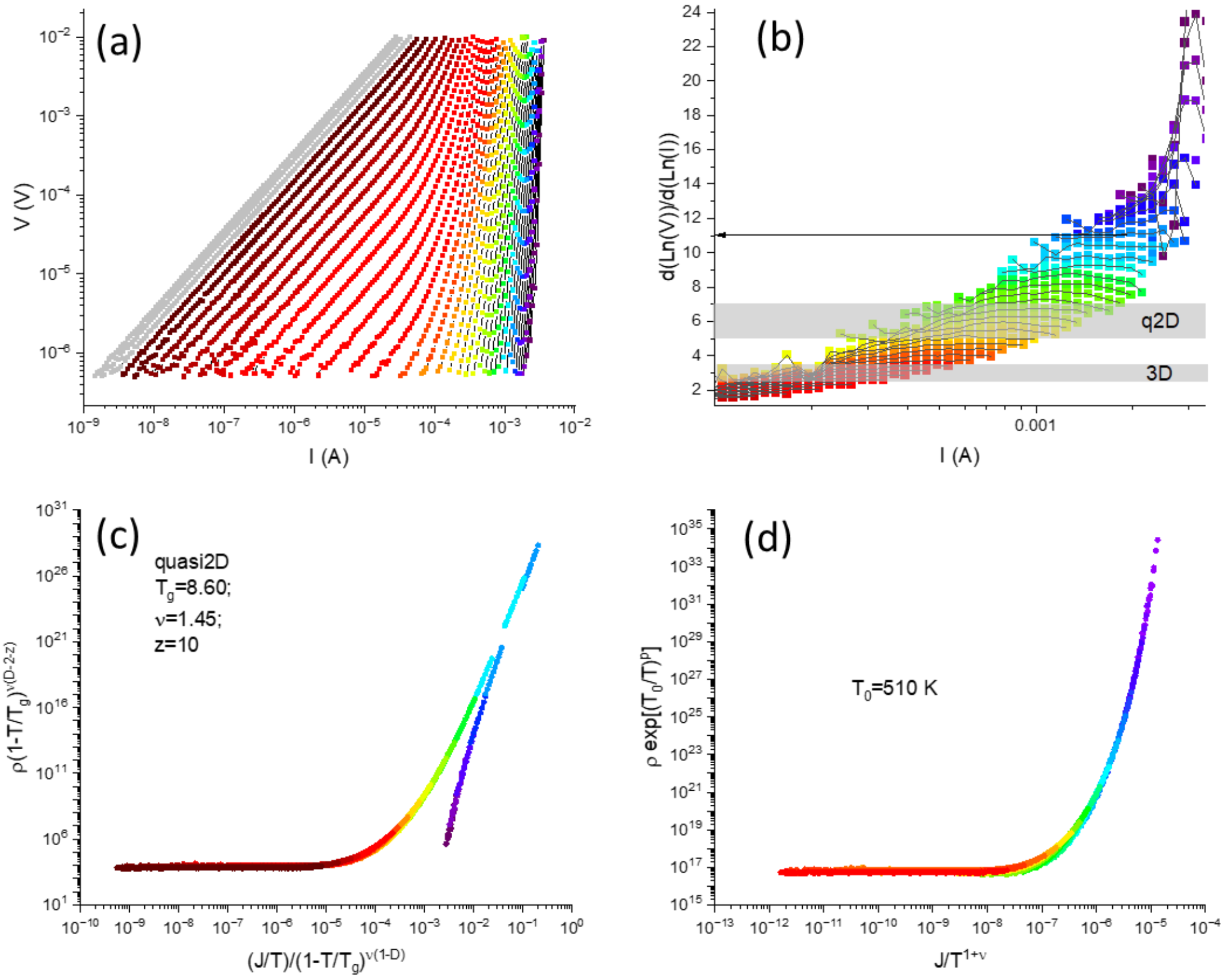

**Figure S5:** The vortex-glass transition in 3D or quasi-2D systems occurs at finite $T_g$, at which $V(I)$ changes from concave upward to concave downward, signaling the onset of a true superconducting state with $\lim_{I\to 0} V/I = 0$. $V(I)$ at $T_g$, known as the critical isotherm, separates $V(I)$ of opposite curvatures and, according to the vortex-glass theory [1–3], follows the scaling $V \propto I^{\alpha+1}$ in the low current limit. Here, $\alpha + 1 = (z - D + 2)/(D - 1) + 1$, with $D = 3$ for three-dimensional or $D = 2$ for quasi-two-dimensional systems, and $z$ is the static critical exponent that should take values $4 < z < 6$ according to the theory [1]. **(a)** Double-logarighmic V(I) isotherms for sample #3 measured under $\mu_0 H$ =0.7 T for 6.5 K $< T <$ 16 K in 0.25 K steps. **(b)** Derivative of the curves shown in (a). In a double log representation, the critical isotherm must show linear behavior in the low-current limit and a slope 5< $\alpha + 1$ <7 for D=2 and 2.5< $\alpha + 1$ <3.5 for D=3 (grey shades in the (b) indicate those ranges). However, in (b), all the isotherms with slope in that range, and many above them, show upward curvature in the low-current limit. In contrast to the behavior observed in Fig. 3 (b), no pair of isotherms of opposite curvature can be identified around a critical isotherm whose slope falls within the expected range. In **(b)**, isotherms with linear response in the low-current limit display a much larger slope ~ 11, corresponding to $z$~10 for D=2 or $z$~19 for D=3. Those values are outside the range predicted by the theory. Whereas a scaling of the $V(I)$ dataset is possible using $z$~10 for D=2 and a finite $T_g$ (see panel **(c)**), [2] earlier work found that such a high value of $z$ indicates a crossover into pure 2D behavior. Indeed, scaling of the dataset with the pure 2D model is fully successful, as shown in panel **(d)**, indicating such crossover from quasi-2D and purely 2D vortex dynamics.

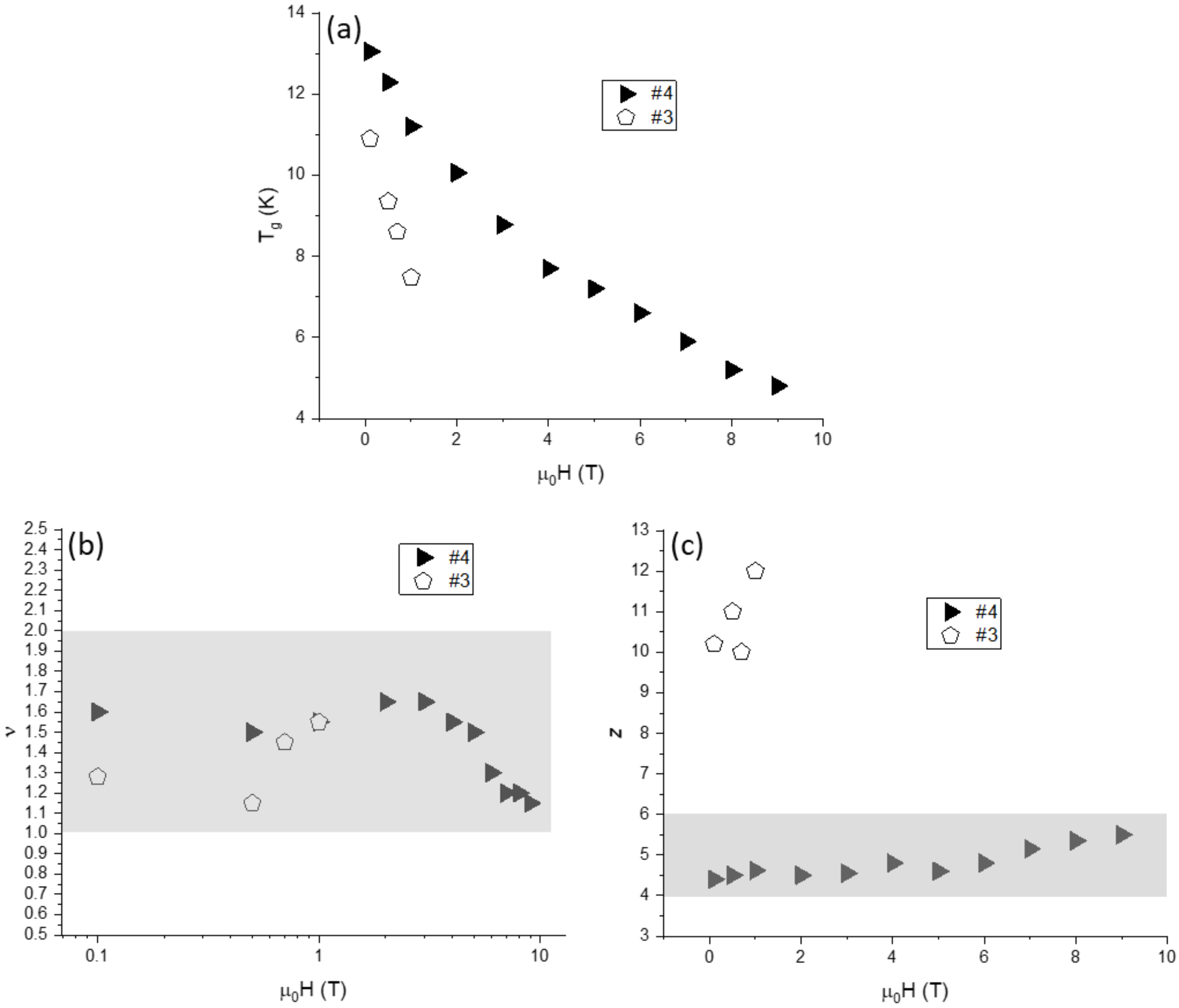

**Figure S6: (a)** Glass transition temperature $T_g$, (b) static and (c) dynamic critical exponent as a function of magnetic field for the samples/applied fields that allow scaling the $V(I)$ datasets according to the quasi-2D vortex glass model. The grey shaded areas indicate the range of values expected from the vortex-glass theory. From sample #4, the quality of the scaling is excellent for all the applied magnetic fields, and the critical exponents lie within the expected range. For sample #3, the dynamic critical exponents (estimated from the analysis of the $V(I)$ derivatives (see supplementary Fig. 4) are very much over the expected range, and increase as the applied field increases (the derivatives do not allow finding $z$ for $\mu_0 H$>1 T). While scaling of the corresponding datasets with a quasi-2D model is successful using the parameters in the figure, the exceedingly high value of $z$, its increase with increasing $H$, and the fact that a scaling with a 2D model is excellent [see supplementary Fig. 4 (d)] indicate that sample #3 is undergoing a crossover into a pure 2D model in the field range 0.1 T< $\mu_0 H$ <1 T, and behaves a pure 2D system beyond that field.

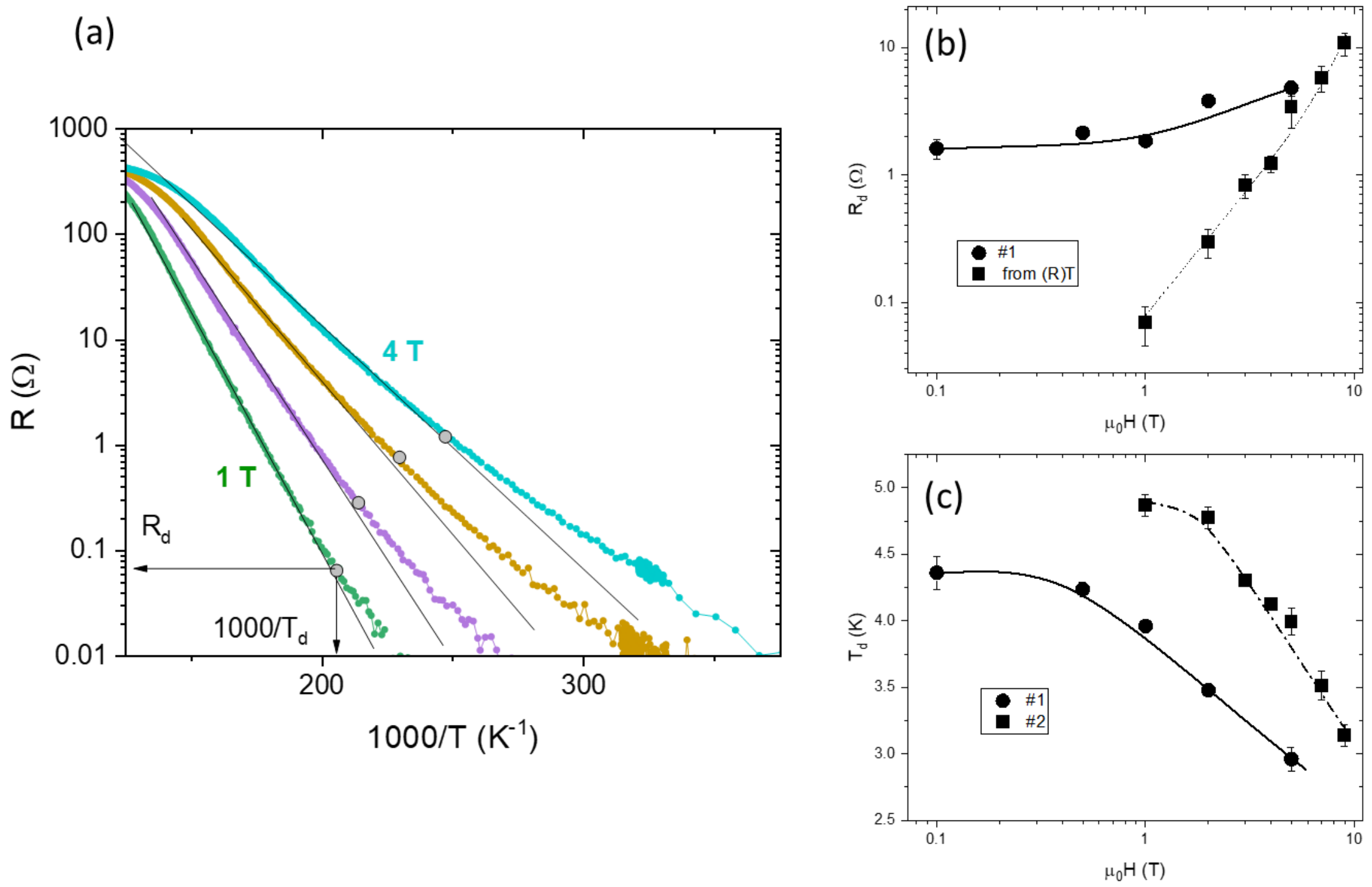

**Figure S7: (a)** Arrhenius plot of the resistive transition under magnetic fields 1T, 2T, 3T and 4T for sample #2, measured with $I$=10 µm. A clear departure from thermally activated behavior is observed below a characteristic temperature $T_d$, where the resistance exceeds the extrapolation of the high-temperature activated trend. **(b)** shows the resistance level $R_d$ and **(c)** $T_d$ at the departure from thermally activated behavior, for samples #1 and #2. Lines are a guide to the eye. The overall behavior is reminiscent of that observed in highly disordered 2D superconducting films [4], and suggests a crossover to a temperature-independent resistance state resulting from the quantum tunneling of vortices. Consistently, this phenomenology is only observed here in samples #1 and #2 from the series, which show a pure 2D vortex-glass transition and the highest degree of disorder, as indicated by their normal-state resistivity.

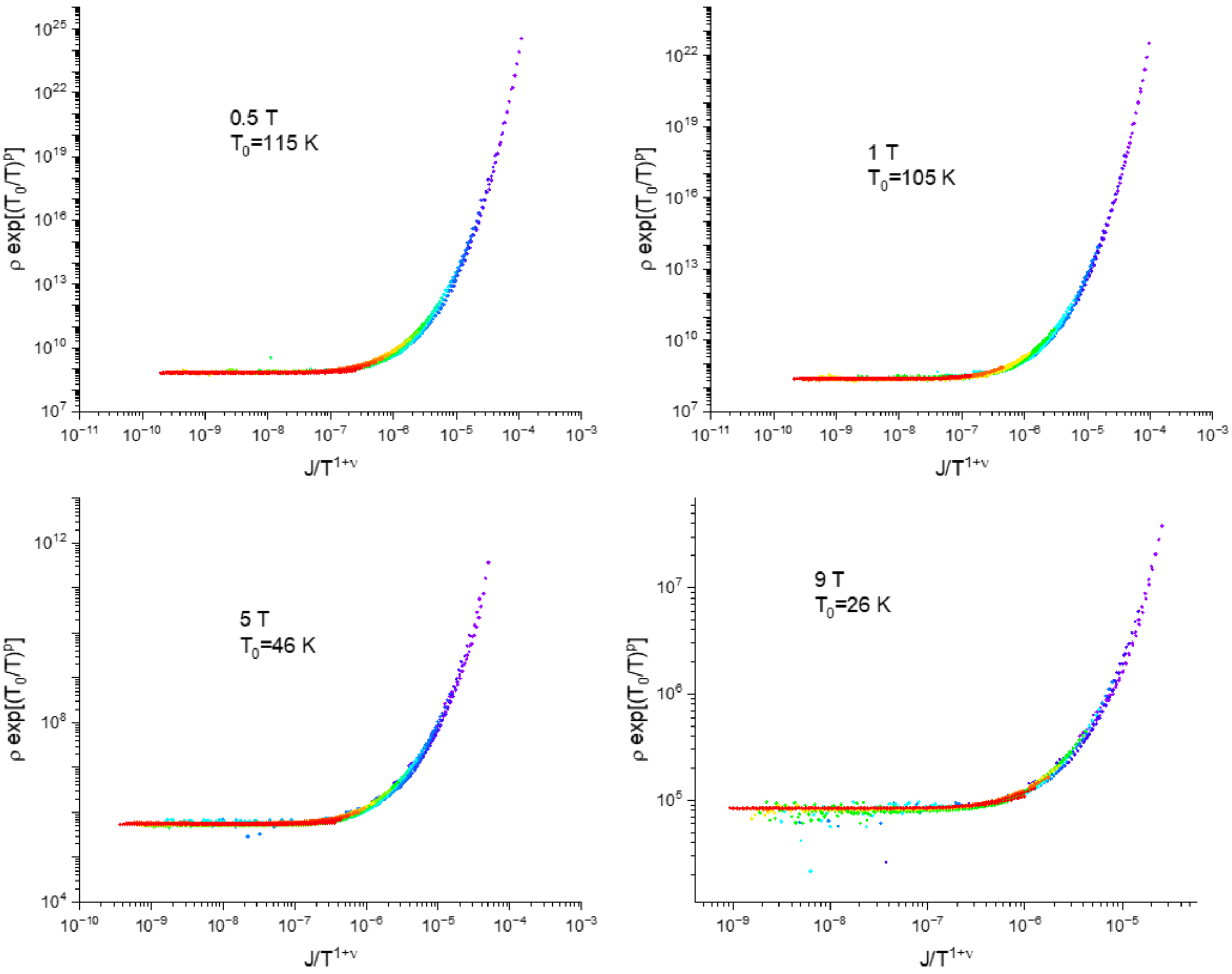

**Figure S8:** For sample #2: scaling of $V(I)$ according with the 2D model for various magnetic fields in the explored range (0.1T to 9 T). Measurements for 0.1 T are shown in Fig. 3. $R_{lin}$ data points for $T < T_d$ (temperature of deviation from activated behavior, see supplementary Fig. 2) cannot be scaled and are excluded. The non-linear (high-current) data points for $T < T_d$ isotherms do collapse onto the master curve.

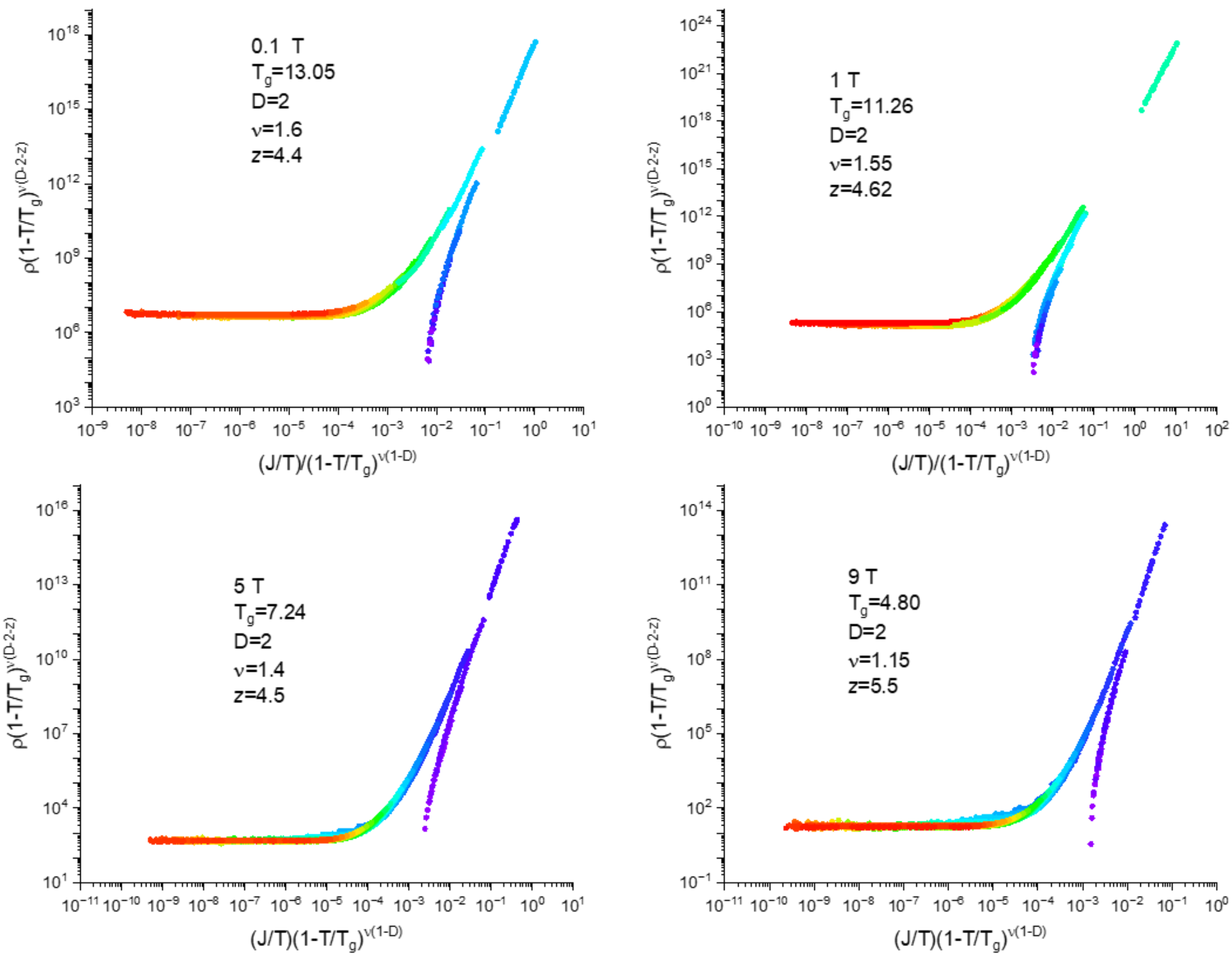

**Figure S9:** For sample #4: scaling of $V(I)$ according to the quasi-2D model for various magnetic fields in the explored range (0.1T to 9 T). Measurements for 4T are shown in Fig. 3.